\documentclass[fleqn,10pt]{wlscirep}
\usepackage{amsmath}
\usepackage{nccmath}
\usepackage{mathtools}
\usepackage{multicol}

\title{Chaotic Dynamics Enhance the Sensitivity of Inner Ear Hair Cells}
\author[1]{Justin Faber}
\author[1,2,*]{Dolores Bozovic}
\date{\today}

\affil[1]{Department of Physics \& Astronomy, University of California, Los Angeles, California 90095, USA}
\affil[2]{California NanoSystems Institute, University of California, Los Angeles, California 90095, USA}
\affil[*]{bozovic@physics.ucla.edu}

\begin{document}

\begin{abstract}
\noindent \textbf{Hair cells of the auditory and vestibular systems are capable of detecting sounds that induce sub-nanometer vibrations of the hair bundle, below the stochastic noise levels of the surrounding fluid.  Hair bundles of certain species are also known to oscillate without external stimulation, indicating the presence of an underlying active mechanism.  We propose that chaotic dynamics enhance the sensitivity and temporal resolution of the hair bundle response, and provide experimental and theoretical evidence for this effect. By varying the viscosity and ionic composition of the surrounding fluid, we are able to modulate the degree of chaos observed in the hair bundle dynamics \textit{in vitro}.  We consistently find that the hair bundle is most sensitive to a stimulus of small amplitude when it is poised in the weakly chaotic regime.  Further, we show that the response time to a force step decreases with increasing levels of chaos.  These results agree well with our numerical simulations of a chaotic Hopf oscillator and suggest that chaos may be responsible for the sensitivity and temporal resolution of hair cells.}  
\end{abstract}
\maketitle

\begin{multicols}{2}


\noindent The auditory system exhibits extraordinary sensitivity and temporal resolution.  It is capable of detecting sounds that induce motion in the {\AA} regime, below that of the stochastic noise levels of the surrounding fluid.\cite{HUDSPETH14}  Humans are able to resolve two stimulus impulses that are temporally separated by only 10 microseconds, where the stimulus waveform is presented simultaneously into both ears.\cite{LESHOWITZ71}  These and other remarkable features of the auditory system are not fully understood, and the physics of hearing remains an active area of research.\cite{REICHENBACH14} 

Mechanical detection is performed by hair cells, specialized sensory cells named after the organelle that protrudes from their apical surface. This organelle, called the hair bundle, consists of rod-like stereovilli that are organized in interconnected rows.  An incoming sound wave pivots the hair bundle, modulating the open probability of the transduction channels that are embedded at the tips of the stereovilli. The mechanical energy of a sound wave is thus transduced into an electrical potential change, due to the influx of ionic current.\cite{LEMASURIER05, VOLLRATH07}  

Hair bundles of several species have been shown to oscillate even in the absence of a stimulus.\cite{BENSER96, MARTIN03}  These spontaneous oscillations are a manifestation of an internal active process, as they violate the fluctuation dissipation theorem.\cite{MARTIN01,HUDSPETH08}  Spontaneous oscillations of the hair bundle hence provide a useful experimental probe for studying the underlying active mechanism \textit{in vitro}.\cite{MARTIN03}  While their role \textit{in vivo} remains to be determined, the presence of spontaneous otoacoustic emissions\cite{KEMP79} suggests that active innate oscillators may be present in the inner ears of intact animals.  Several numerical studies have applied dynamical systems models to demonstrate that spontaneous motility could produce otoacoustic emissions.\cite{VILFAN08, FRUTH14} 

Dynamics of the auditory response have been modeled using the normal form equation for Hopf bifurcations.\cite{EGUILUZ00, KERN03}  This simple differential equation accounts for many experimentally observed phenomena, including the sensitivity and frequency selectivity of hearing exhibited by many species. To reproduce the empirically measured sensitivity, the models have assumed that the system is poised in close proximity to the Hopf bifurcation. This assumption raises the question of how the biological system achieves and then maintains such fine-tuning of the parameters. To circumvent this issue, some models include a dynamic feedback equation responsible for automatically tuning the control parameter towards or away from criticality.\cite{CAMALET00, SHLOMOVITZ13}  Other studies proposed that the inclusion of a homeostatic equation can broaden the parameter regime of extreme sensitivity, frequency selectivity, and compressive nonlinearity.\cite{MILEWSKI17}  A second issue with proximity to criticality is the phenomenon of critical slowing down: near the bifurcation, a system would exhibit a slow response, which seems inconsistent with the high temporal resolution that characterizes hearing. This second objection is not resolved by the inclusion of homeostasis or feedback.  

We propose that the system is poised in the oscillatory state, not in the immediate vicinity of the Hopf bifurcation. We focus our theoretical and experimental studies on that regime, as it is consistent with the occurrence of spontaneous otoacoustic emissions \textit{in vivo}, a phenomenon that is ubiquitous across vertebrate species. In a prior study, we demonstrated experimentally that spontaneous oscillations exhibit chaotic dynamics.\cite{FABER18}  Further, we showed theoretically that the Hopf oscillator exhibits enhanced temporal resolution and sensitivity when poised in the unstable regime where noise induces chaos.\cite{FABER19}  As chaotic oscillators are a subclass of nonlinear systems that exhibit extreme sensitivity to initial conditions,\cite{ANISHCHENKO07} we propose that chaos leads to both high sensitivity and rapid response to mechanical perturbation that characterizes hair bundle dynamics.

In the current work, we show that, for a wide range of parameter conditions, additive noise induces chaotic dynamics.  We use a simple theoretical model to demonstrate that extreme sensitivity and rapid response time of the chaotic system both occur over a wide parameter range, without the need for a feedback equation.  This alternative view could explain how the performance of the hair cells would be robust to noise as well as to changes in parameters, and how it can achieve extreme sensitivity in the oscillatory regime, from which otoacoustic emissions can originate. In the numerical model, we vary the degree of chaos and show that the sensitivity to different stimulus waveforms is enhanced and the response time reduced, as the degree of chaos is increased.  We propose to view the hair cell not just as a mechanical resonator, but as an information processor that extracts selective information from its acoustic environment.  We then show that the amount of information extracted by the active oscillator from an imposed stimulus waveform is maximized in the weakly chaotic regime.  We verify our theoretical predictions by experiments performed on \textit{in vitro} preparations of the bullfrog sacculus. By varying the viscosity and the ionic conditions of the fluid in which the hair bundles are immersed, we modulate the chaoticity of their dynamics, and we measure the sensitivity, information transfer, and temporal response to various imposed signals. We find consistent experimental agreement with all of the theoretical predictions of the model.  We therefore propose that the instabilities giving rise to chaotic dynamics enhance the sensitivity and temporal resolution of the auditory and vestibular hair cells.


\begin{figure*}[t!]
\includegraphics[width=\textwidth]{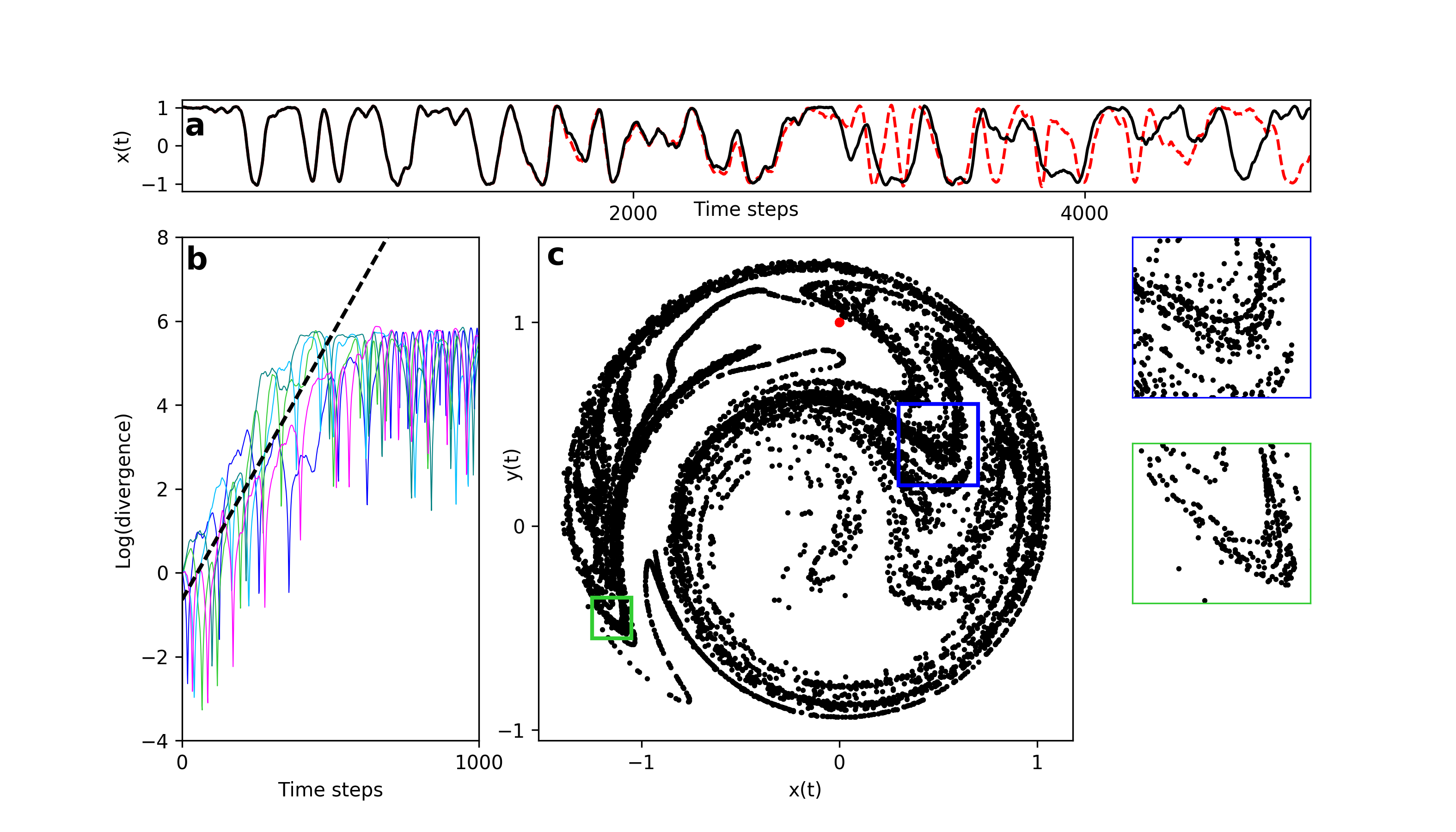}
\caption{(\textbf{a})  The divergence of two neighboring solutions to equation (\ref{eq:Hopf}). The two time-dependent solutions are depicted with black (solid) and red (dashed) lines.  (\textbf{b}) The natural logarithm of the average separation of neighboring trajectories.  Each of the five colors represents an average of 200 pairs of neighboring trajectories, taken with different initial conditions and realizations of common noise.  The dashed line represents the linear fit to all of the data within the first 400 time steps.  (\textbf{c})  The spreading of trajectories throughout the phase space.  $10^4$ initial conditions were randomly selected in the same neighborhood (red points).  After 500 time steps,  these solutions spread across the phase space to reveal the fractal structure of the attractor (black points).  Side panels show zoomed-in sections corresponding to the colored squares in the main figure.}
\label{Fig1}
\end{figure*}

\begin{figure*}[t!]
\includegraphics[width=\textwidth]{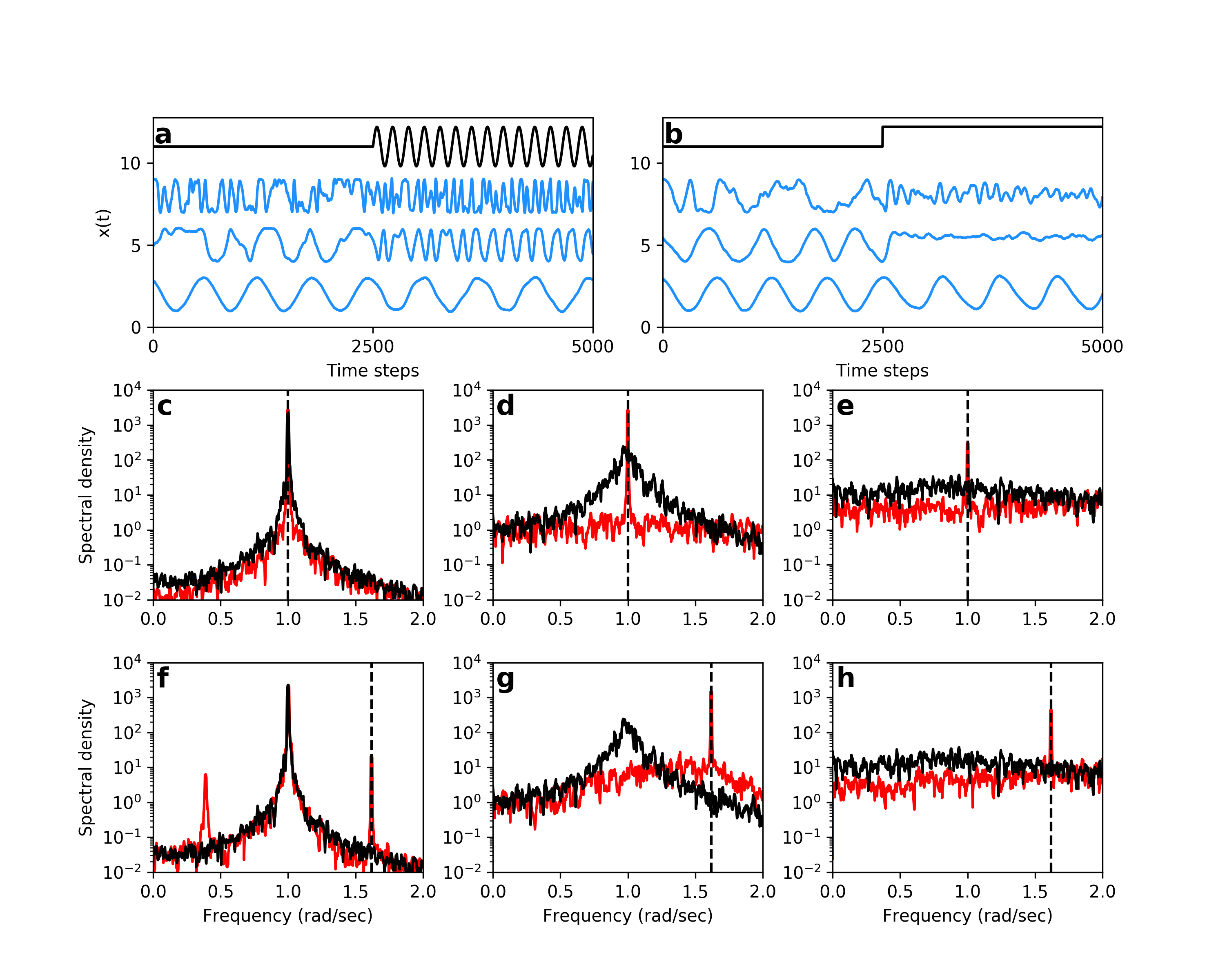}
\caption{(\textbf{a-b}) Time-domain responses to (off-resonance) sinusoidal and step stimulus, respectively.  The top, black traces show the stimulus waveform.  Bottom, middle, and top blue traces represent responses of a system with no chaos, weak chaos, and strong chaos, respectively.  The degree of chaos was modulated by varying $\beta$ and $\omega_0$, while keeping the natural frequency fixed at $\Omega_0 = 1$.  (\textbf{c-e})  Power spectral density of the response to on-resonance sinusoidal stimulus ($\omega_{stim} = \Omega_0 = 1$, as indicated by the vertical, dashed lines) for systems exhibiting no chaos (c), weak chaos (d), and strong chaos (e).  Red and black curves represent the spectral density in the presence and absence of the stimulus, respectively.  (\textbf{f-h})  Power spectral density of the response to off-resonance sinusoidal stimulus ($\omega_{stim} = \frac{1+\sqrt{5}}{2}\Omega_0$, as indicated by the vertical, dashed lines) for systems exhibiting no chaos (f), weak chaos (g), and strong chaos (h).  The stimulus frequency was set to the golden ratio with respect to the natural frequency to avoid mode locking.  Red and black curves represent the spectral density in the presence and absence of the stimulus, respectively.}
\label{Fig2}
\end{figure*}

\begin{figure*}[t!]
\includegraphics[width=\textwidth]{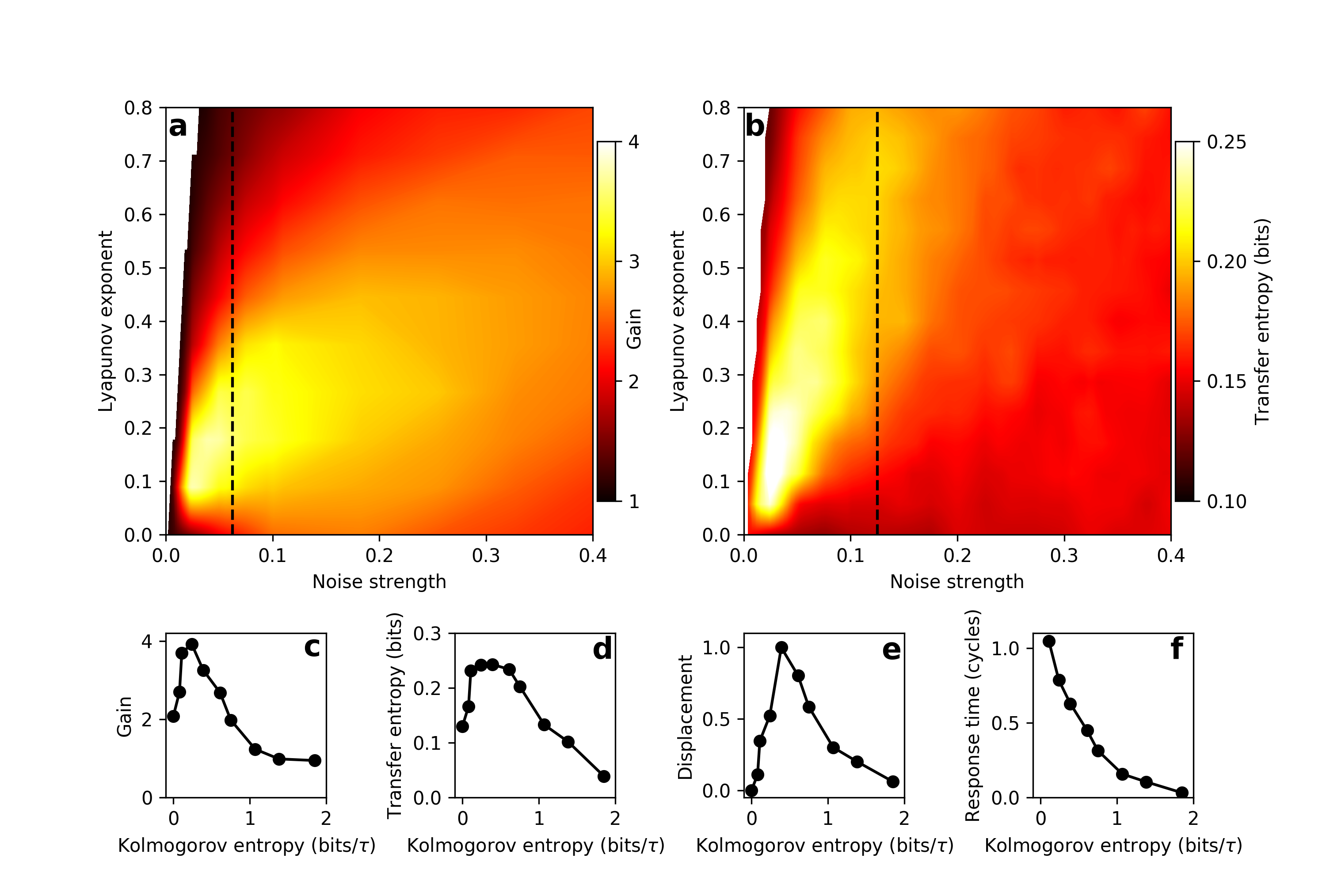}
\caption{(\textbf{a}) Phase-locked amplitude gain for on-resonance, sinusoidal stimulus.  (\textbf{b})  Transfer entropy from burst noise stimulus to response of the Hopf oscillator.  For (a-b), $\beta$, $\omega_0$, and $D$, were varied, resulting in a range of Lyapunov exponents.  All other parameters were fixed ($\mu = \alpha = \Omega_0 = 1$).  Color was generated by linearly interpolating a grid of 21 Lyapunov exponent values and 29 noise strengths.  The stimulus amplitude was set to $0.5$ for both panels.  The vertical dashed lines indicate the points where the signal-to-noise ratio is $1$, as defined by the ratio of the signal power to the noise power.  (\textbf{c}) Phase-locked amplitude gain for on-resonance, sinusoidal stimulus.  (\textbf{b})  Transfer entropy from burst noise stimulus to response of the Hopf oscillator.  (\textbf{c}) Mean displacement in response to a step stimulus.  (\textbf{d}) Response time to the step stimulus, calculated by taking the decay time of an exponential fit to the mean response.  For (c-f), the Kolmogorov entropy was modulated by varying $\beta$ and $\omega_0$.  All other parameters were fixed ($\mu = \alpha = \Omega_0 = 1$, $D = 0.05$).}
\label{Fig3}
\end{figure*}

\begin{figure*}[t!]
\includegraphics[width=\textwidth]{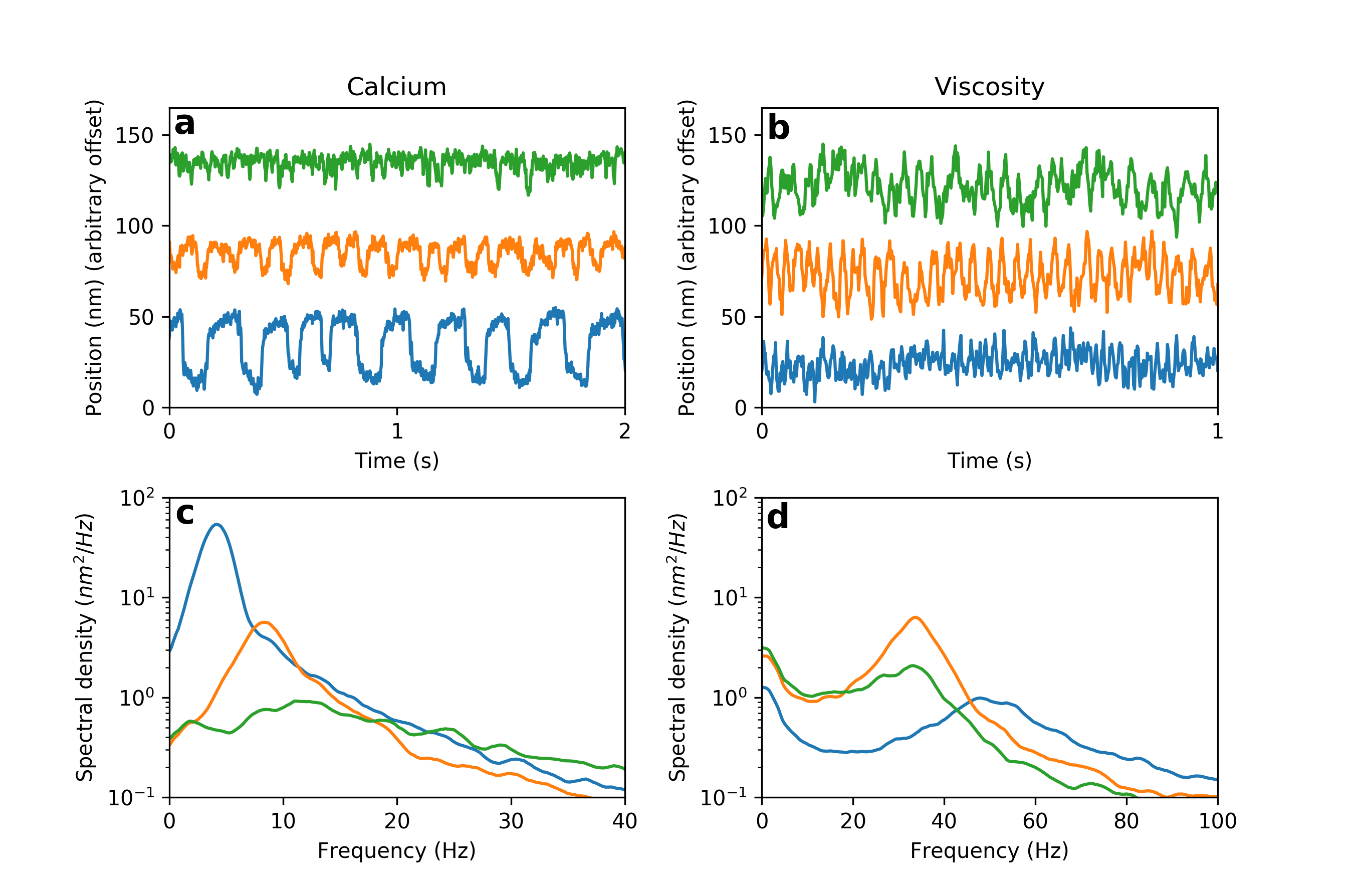}
\caption{(\textbf{a})  Spontaneous oscillations of a hair bundle under various calcium concentrations of the endolymph.  From bottom to top: 100 $\mu M$ (low calcium), 250 $\mu M$ (natural calcium), and 325 $\mu M$ (high calcium).  (\textbf{b})  Spontaneous oscillations of a hair bundle with various endolymph viscosities.  From bottom to top: 0, 70, and 100 $\frac{mg}{ml}$ of Dextran 500.  (\textbf{c}) Power spectral density of the traces in (a).  (\textbf{d}) Power spectral density of the traces in (b).}
\label{Fig4}
\end{figure*}

\begin{figure*}[t!]
\includegraphics[width=\textwidth]{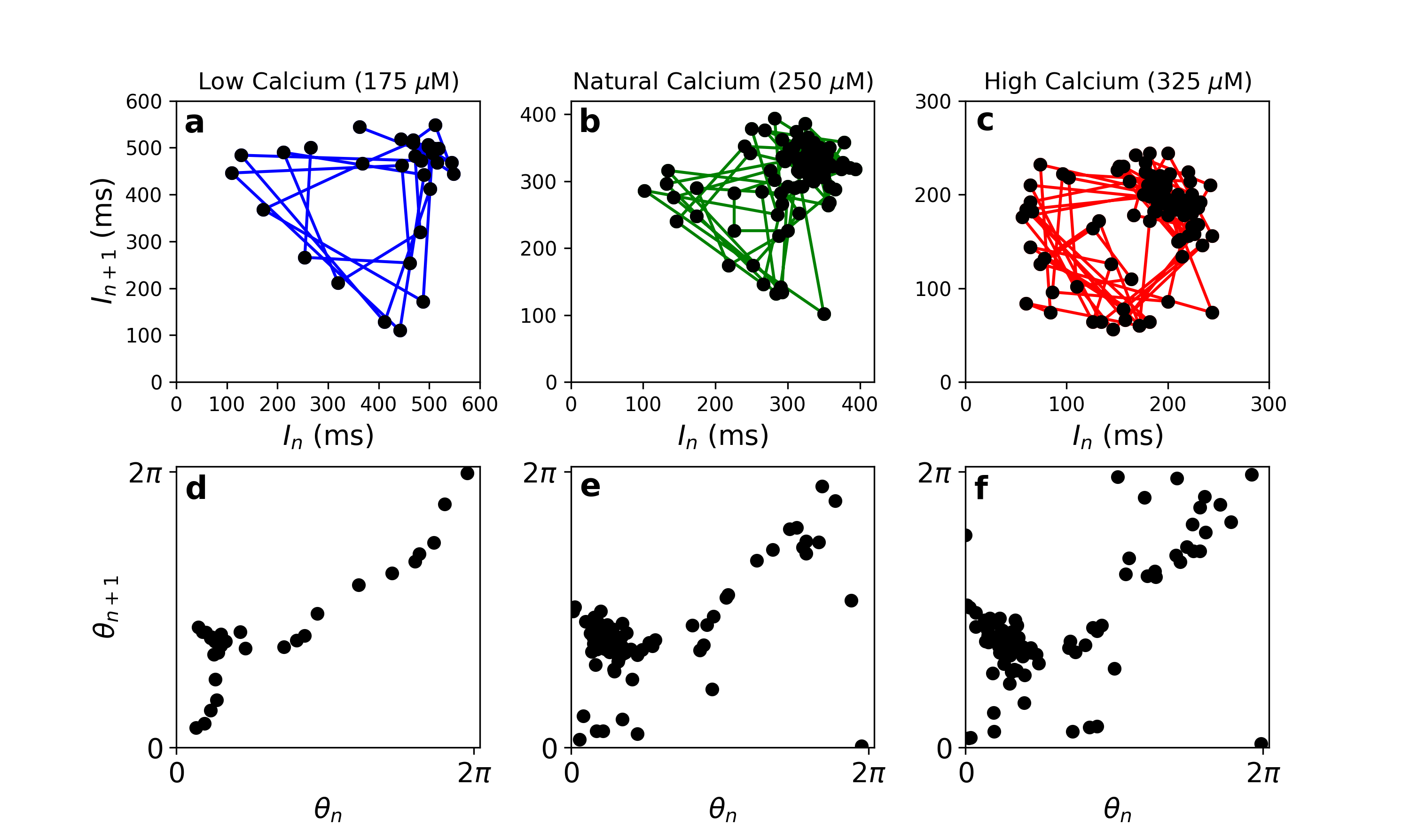}
\caption{(\textbf{a}-\textbf{c})  Poincar\'e maps constructed from the time intervals between the steepest rising flanks of consecutive hair bundle oscillations under low ($175 \mu M$), natural ($250 \mu M$), and high ($325 \mu M$) calcium concentrations of the endolymph, during presentation of off-resonance stimulus. (\textbf{d})  Circle map corresponding to the low-calcium conditions.  The monotonic function suggests the absence of chaos.  (\textbf{e}-\textbf{f}) Circle maps corresponding to the natural- and high-calcium conditions, respectively.  The absence of a monotonic function indicates the presence of chaos.  We use Spearman's rank correlation coefficient to test for monotonicity of the circle maps.  Under low-, natural-, and high-calcium conditions, Spearman's coefficient is 0.60 $\pm$ 0.01, 0.14 $\pm$ 0.01, and 0.30 $\pm$ 0.02, respectively.  Uncertainties represent 1 standard deviation from a sample of 100 bootstraps.}
\label{Fig5}
\end{figure*}

\begin{figure*}[t!]
\includegraphics[width=\textwidth]{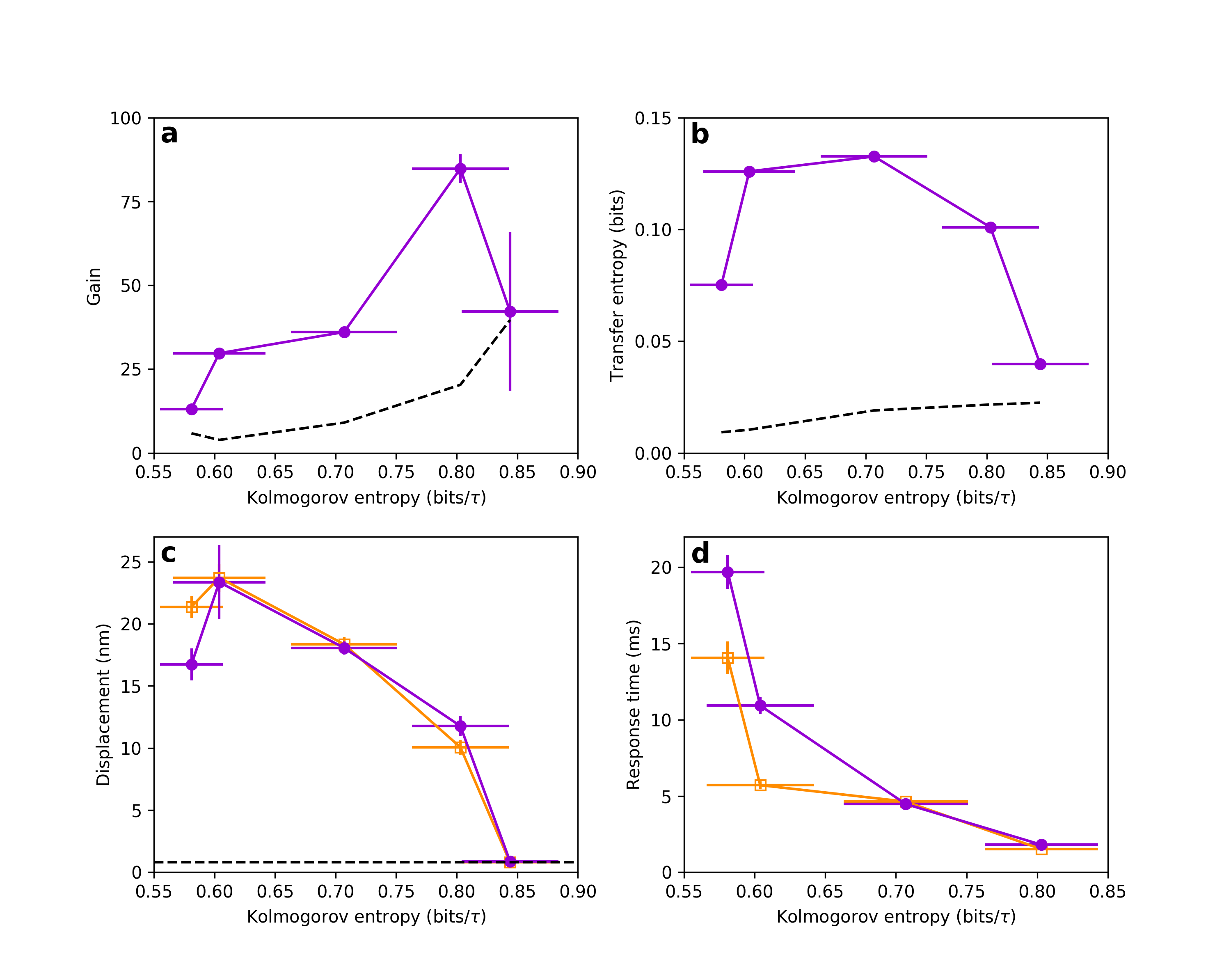}
\caption{(\textbf{a}) Phase-locked amplitude gain for 2 pN sinusoidal stimulus, presented at the natural frequency.  Data points and error bars on this measure represent the mean and standard deviation from 100 bootstraps. The noise floor (dashed curve) was calculated by treating a segment of the spontaneous oscillation recording as if a stimulus were present and calculating the gain.  This curve represents the mean plus one standard deviation from 100 bootstraps.  (\textbf{b}) Transfer entropy from burst noise stimulus to hair bundle response.  Data points and error bars represent the mean and standard deviation obtained from 100 bootstraps.  The noise floor (dashed curve) was determined by calculating the transfer entropy in the reverse direction (response to stimulus).  This curve represents the mean plus one standard deviation from 100 bootstraps.  (\textbf{c})  Average displacement induced on the hair bundle from the step stimulus, averaged over $\sim200$ square waves.  Data points and error bars represent the mean and standard deviation of the response plateau.  Orange-open and purple-filled data points represent averages over steps in the positive (channel open) and negative (channel closed) directions, respectively.  The noise floor is represented by the dashed line.  (\textbf{d})  Response time to step stimulus, characterized by fitting the mean response to an exponential and extracting the decay time.  Error bars represent the standard deviation of the residual associated with the exponential fit. All measurements were performed on the same cell.}
\label{Fig6}
\end{figure*}

\section*{Numerical Model}

We use the normal form equation for the supercritical Hopf bifurcation with additive Gaussian white noise, $\eta_z(t)$:

\begin{ceqn}
\begin{align}
\frac{dz(t)}{dt} = (\mu + i\omega_0)z(t) - (\alpha + i\beta)|z(t)|^2z(t) + \eta_z(t),
\label{eq:Hopf}
\end{align}
\end{ceqn}

\noindent where

\begin{ceqn}
\begin{align}
z(t) = x(t) + iy(t),
\end{align}
\end{ceqn} 

\begin{ceqn}
\begin{align}
\eta_z(t) = \eta_x(t) + i\eta_y(t),
\end{align}
\end{ceqn}

\begin{ceqn}
\begin{align}
\langle \eta_x(t)\eta_x(t') \rangle  =  \langle \eta_y(t)\eta_y(t') \rangle = 2D \delta (t-t'),
\end{align}
\end{ceqn}

\noindent and

\begin{ceqn}
\begin{align}
\langle \eta_x(t)\eta_y(t') \rangle  = 0.
\end{align}
\end{ceqn}

\noindent Here, $x(t)$ represents the bundle position, while $y(t)$ reflects internal parameters of the bundle and is not assigned a specific measurable quantity.  $\mu$ represents the control parameter of the system, with positive values yielding a limit cycle and negative values resulting in a stable fixed point.  The autonomous angular frequency of this system in the absence of noise is  $\Omega_0 = \omega_0 - \beta r_0^2$, where $r_0 = \sqrt{\frac{\mu}{\alpha}}$ is the radius of the limit cycle.  Thus, $\omega_0$ represents the angular frequency at the Hopf bifurcation ($\mu=0$).  $\alpha$ and $\beta$ characterize the cubic nonlinearity of the system, while $D$ represents the noise strength.  Auditory and vestibular stimuli induce lateral deflections on the hair bundle, so we consider forces in the $\hat{x}$ direction, which coincides with the direction of spontaneous oscillation.

A characteristic feature of chaotic systems is that neighboring solutions to the differential equations diverge exponentially with time: $\lvert \Delta z(t) \rvert \approx e^{\lambda t}\lvert \Delta z(0) \rvert$, where $\Delta z(t)$ is the separation between two neighboring trajectories in phase space, and $\lambda$ is the Lyapunov exponent.\cite{ANISHCHENKO07}  Thus, small perturbations to a chaotic system will have a drastic impact on future dynamics.  The Lyapunov exponent characterizes how quickly the perturbations grow and serves as a measure of the degree of chaos.  A stable fixed point is characterized by a negative Lyapunov exponent, as neighboring trajectories converge to the same location.  A limit cycle has a Lyapunov exponent equal to zero, since a perturbation tangential to the direction of motion neither grows nor shrinks.  A positive value of the Lyapunov exponent indicates chaotic dynamics, with larger values corresponding to more irregular behavior and weaker predictability.  

In the presence of stochastic processes, the Lyapunov exponent is calculated by measuring the exponential rate of divergence of two neighboring trajectories, subjected to identical realizations of noise (i.e. common noise).\cite{GOLDOBIN05, GOLDOBIN06, NEIMAN11}  This has been observed in other 2-dimensional systems\cite{GAN06} and is commonly referred to as noise-induced chaos.  In Fig. \ref{Fig1}, we demonstrate that the additive noise to equation \ref{eq:Hopf} induces chaotic dynamics, as it causes solutions to diverge exponentially. Further, the attractor exhibits a fractal structure, which is ubiquitous to chaotic systems.\cite{Strogatz94}  The simplicity of this model allows us to use an analytic approximation of the Lyapunov exponent, applying Fokker-Planck theory (see Methods):

\begin{ceqn}
\begin{align}
\lambda \approx \frac{\lvert\beta\rvert D}{\mu}
\label{eq:LyapExp}.
\end{align}
\end{ceqn}

\noindent Without loss of generality, we set $\alpha = \Omega_0 = 1$, scaling the units of length and time.  The remaining three parameters, $\mu$, $\beta$, and $D$ are used to modulate the Lyapunov exponent.

The Lyapunov exponent provides the simplest measure of chaos for numerical models.  However, the Kolmogorov entropy (K-entropy) constitutes a more appropriate measure of chaoticity for experimental data sets, which inherently contain measurement noise and are more limited in duration than typical numerical simulations.\cite{GRASSBERGER83}  Similar to the Lyapunov exponent, the K-entropy measures the divergence rate of neighboring trajectories.\cite{KOLMOGOROV58}  Specifically, K-entropy quantifies the rate at which phase space information is lost due to an expansion of measurement uncertainty.  K-entropy is zero for non-chaotic systems, non-zero for chaotic systems, and infinity for white noise.  We therefore use the K-entropy as our measure of chaoticity when making direct comparisons between theory and experiment.

\section*{Theoretical Results}

We next explore the effects of chaos and noise on the system's sensitivity to a weak stimulus.  The oscillator demonstrates higher responsiveness to sinusoidal and square wave stimuli when it is poised in the weakly chaotic regime (Fig. \ref{Fig2}).  We note that the traditional linear response function is not an appropriate measure of sensitivity in the oscillatory regime, as it would yield infinite sensitivity at the natural frequency, in the limit of vanishing stimulus amplitude and noise.  Instead, we use the amplitude gain at the frequency of the applied signal as a measure of the system's responsiveness. \cite{VILFAN08, FABER18}  We define $gain = \frac{\lvert\tilde{x}(\omega_{stim})\lvert}{\rvert\tilde{x}_0(\omega_{stim})\lvert}$, where $\tilde{x}(\omega_{stim})$ and $\tilde{x}_0(\omega_{stim})$ are the Fourier components at the stimulus frequency in the presence and absence of stimulus, respectively.  Fig. \ref{Fig2} shows sample traces that can be used to calculate the gain.  Notice that the weakly chaotic regime is easily entrained to off-resonance stimulus frequencies.  We consistently find that the chaotic regime is more sensitive than the stable limit cycle regime.  Upon a further increase in the chaoticity of the system, reflected by a higher Lyapunov exponent, the sensitivity deteriorates.  

We measure the gain of the system in response to an on-resonance sinusoidal stimulus for a wide range of Lyapunov exponent values and noise strengths (Fig. \ref{Fig3}a).  We find that the system is most responsive in the weakly chaotic regime (low, but nonzero Lyapunov exponent).  Importantly, this is true even at high levels of additive noise (signal-to-noise ratio $<$ 1).  Further, we see similar results for off-resonance stimuli (see supplemental material, Fig. S2).

Next, we explore the effects of chaos and noise on the amount of signal information captured by the detector (transfer entropy\cite{SCHREIBER00}).  In contrast to measures such as mutual information, transfer entropy explicitly identifies the direction of information flow.  For continuous stimulus and response signals, calculation of transfer entropy requires discretizing the range of the signals and assigning a state for each bin (see Methods).  We therefore use a square wave stimulus with stochastic variation of the period (burst noise or random telegraph noise).\cite{FABER18}  Such stochastic signals constantly produce new information, and the transfer entropy measures how much of this new information is captured by the detector.  The square wave intervals are randomly generated from a flat distribution that spans two octaves on either side of the natural frequency ($\frac{1}{4}\Omega_{0}$ to $4\Omega_{0}$).  We measure the transfer entropy over a wide range of Lyapunov exponent values and noise strengths and find results consistent with the measurements of the gain (Fig. \ref{Fig3}b).  We also show that consistent results are obtained when $\mu$ is used to modulate the Lyapunov exponent in place of $\beta$ (see supplemental material, Fig. S3).  The weakly chaotic regime is optimal for extracting information from the signal.  This feature persists at high levels of additive Gaussian white noise (signal-to-noise ratio $<$ 1).

As an additional test of the system's response, we also measure the mean displacement induced by a step stimulus.  We apply a step stimulus to the system and average over many different initial conditions and realizations of noise.  We then compute the difference between the averaged curves before the onset of the step stimulus and after the system settles to the new steady state.  Both the transfer entropy and the  mean response amplitude show a local maximum as the degree of chaos is varied (Fig. \ref{Fig3}d-e), consistent with the sensitivity observed in response to sinusoidal stimuli.

Finally, we measure the characteristic time of the exponential rise of the averaged response to the step-function forcing.  We use this response time to characterize the temporal resolution of the system.  The response time decreases with increasing levels of chaos (Fig. \ref{Fig3}f). Increasing chaoticity therefore allows the system to exhibit a faster response to an external perturbation.

\section*{Experimental Results}
   
We use two experimental parameters to modulate the degree of chaos exhibited by oscillatory hair cells \textit{in vitro}.  The first parameter varied is the calcium concentration of the endolymph solution, which has been shown to affect the dynamics of the adaptive mechanisms within the hair bundle.\cite{EATOCK00}  Varying the calcium concentration alters the spontaneous oscillation profile,\cite{MARTIN03} with higher concentrations resulting in more irregular dynamics (Fig. \ref{Fig4}a,c).  As expected, increasing the calcium concentration increases the degree of chaos (see supplemental material, Fig. S4).  

The second experimental parameter we vary is the viscosity of the endolymph solution.  It has recently been shown that increasing the viscosity suprisingly increases the regularity of the spontaneous oscillations.\cite{MARTIN18}  Once the viscosity is increased beyond about five times its natural value, the regularity of the spontaneous oscillations then decreases (Fig. \ref{Fig4}b,d).  We vary both of these parameters in our experiments in order to densely sample the parameter space of this chaotic oscillator.

While the variations described above yield visible differences in the regularity of the oscillations, rigorous mathematical tests are required to establish whether the active motility is chaotic or not. One reliable method for determining the presence of chaos in a system is provided by observing the type of transition it undergoes as it phase-locks to an external signal.  If the external signal induces a torus-breakdown transition, this feature is an indicator of chaos.  We hence constructed Poincar\'e maps of hair bundle oscillations driven by sinusoidal mechanical perturbations of increasing amplitude,\cite{FABER18} following methods developed earlier.\cite{HEGGER97}  We determine the discrete time series, $[I_n]$, where each element reflects the time interval between the steepest rising flanks of consecutive hair bundle oscillations.  We then plot the $n^{th}$ versus the $(n+1)^{th}$ point of the series to obtain the Poincar\'e map (Fig. \ref{Fig5}a-c).  For low-amplitude, off-resonance stimuli, the points comprise a ring structure, revealing a cross-section of the underlying torus, and are indicative of quasiperiodic dynamics.  Chaotic dynamics arise when this ring structure loses smoothness.\cite{FENICHEL71, ARONSON82, SHILNIKOV03}  To check for smoothness of a Poincar\'e map, we determine the series of angles that each point makes with the abscissa and construct a circle map, $\theta_{n+1} = f(\theta_n)$ (Fig. \ref{Fig5}d-f).  When the surface of the torus is smooth, the map $f$ is a monotonically increasing function.  When chaos arises, the torus loses smoothness, and the map $f$ loses monotonicity and may cease to be a function at all.

Under the natural conditions of the hair cell, we consistently find that off-resonance stimulus induces the torus-breakdown transition to/from chaos. Likewise, under high-calcium conditions, the circle map shows the absence of smoothness indicative of a chaotic system. However, when the hair cell is immersed in a low-calcium endolymph solution, the torus breakdown transition vanishes, and the circle map becomes a function (Fig. \ref{Fig5}d).  This finding suggests that spontaneous oscillations in low-calcium solution are non-chaotic, consistent with the observation of their regularity, reflected in a higher quality factor ($Q > 3$).  

In addition to the Poincar\'e maps, we quantify the degree of chaos in the active dynamics of hair cells by estimating their Kolmogorov entropy, following techniques previously developed for the analysis of experimental records (see Methods).\cite{GRASSBERGER83}  We note that this measure is useful for characterization of the degree of chaos in a system, once other methods have confirmed its presence.  Any amount of measurement noise imposes a noise floor on the K-entropy, and hence, even the most regular spontaneous oscillations yield a small, positive K-entropy. An independent method must therefore be used to identify the crossover from stable to chaotic dynamics, and thus determine the effective noise floor on K-entropy. The circle maps described above indicate a non-chaotic state under low-calcium conditions, with the corresponding K-entropy that is small and positive, at $\sim 0.5$ bits/$\tau$. We hence use this value of K-entropy as an experimental estimate of the noise floor.  

To obtain different levels of chaos in bundle dynamics, we immersed the preparations in solutions of different combinations of calcium concentration and viscosity.  At each experimental condition, we first record the innate oscillations of hair bundles, followed by measurements of their response to sinusoidal stimuli, presented at several fixed frequencies selected to yield both on- and off-resonance responses. Subsequently, we present burst noise, with square waves of various duration, selected from a random distribution (see Methods). Measurements of the response to sinusoidal stimuli allow us to extract the phase-locked amplitude gain as an estimate of the mechanical sensitivity of the system. The burst noise yields the measure of the transfer entropy, as well as that of the mean displacement and response time of the bundle. The same methods are used to analyze the experimental records as those obtained from numerical simulations in the prior section. 

Our findings consistently show optimal sensitivity and information gain in the weakly chaotic regime, as shown in  Fig. \ref{Fig6} and in the supplemental material (Fig. S5).  Upon increasing the degree of chaos, we find that the gain increases by at least 2-fold (even 5-fold for some cells).  The information extracted from the burst noise stimulus also increases by more than 2-fold as the level of chaos is increased.  Furthermore, increasing chaoticity yields a more rapid response time (Fig. \ref{Fig6}d and S6), indicating a higher temporal resolution.  Within the range of experimentally accessible levels of chaos, we see the response time reduce by about 5-fold (3-fold if measured in terms of the natural period of the hair bundle, see Fig. S6g-i).

\section*{Discussion}

The auditory and vestibular systems have provided a powerful experimental testing ground for concepts in nonlinear dynamics,\cite{EGUILUZ00, CAMALET00} nonequilibrium thermodynamics,\cite{DINIS12} and condensed matter theory.\cite{RISLER04}  Some of the long-standing open questions in this field pertain to how hair cells can reliably transform a sound wave into a neural spike train with such sensitivity, frequency selectivity, and temporal resolution.  Most theoretical studies of hair cell detection have focused on the stable limit cycle regime or on the interface between a stable limit cycle and a stable fixed point.  Using the simplest model of hair cell dynamics, we have identified a chaotic state that has greater sensitivity to both sinusoidal and step-function stimuli than either of these traditional regimes in the presence of noise.  Further, we have shown that this chaotic regime extracts more information from its acoustic environment and achieves greater temporal resolution, all while maintaining robustness to additive noise.

Chaos is typically considered a harmful element to dynamical systems.  For example, a chaotic heartbeat is an indicator of cardiac fibrillation.\cite{GARFINKEL97, KIM97}  Chaos may also be responsible for the anti-reliability of neurons.\cite{GOLDOBIN06, STOOP04}  However, in the present work, we have demonstrated that chaos is beneficial to sensory detection by hair cells.  The dynamic state of a chaotic system depends sensitively on its initial conditions, and hence a small perturbation can result in a drastic change in the subsequent dynamics.  We speculate that evolution has exploited this feature of chaos to enable hair cells to achieve sensitivity to displacements in the {\AA} regime. Furthermore, auditory detection relies on high temporal resolution, in order to enable accurate spatial localization of a sound. Our results, obtained both theoretically and experimentally, indicate that a chaotic system exhibits faster response times than one poised in the stable regime. This is again consistent with the general nature of chaotic systems, which show exponential divergence of trajectories in response to a perturbation. We propose that this regime provides an attractive alternative to proximity to the Hopf bifurcation, which achieves high sensitivity, but at the price of critical slowing down.  

As most biological systems are nonlinear and contain many degrees of freedom, chaos is likely to be a ubiquitous feature of their dynamics.  We speculate that many other systems in biology, beyond those currently known, may exhibit chaotic dynamics. In particular, sensory systems that are responsible for detection of external signals may have evolved to harness the power of these instabilities.  In the present work, we explored the effects of chaos on the sensitivity of an individual hair cell, and demonstrated that it enhances its responsiveness. Our future work entails exploring the effects of chaos on the sensitivity of detection in systems of coupled hair cells.

\bibliography{Bibliography}

\subsection*{Data Availability}
The data supporting the findings of this study are available within the article and its supplementary material file.  Raw datasets generated during the current study are available from the corresponding author on reasonable request.

\subsection*{Acknowledgements}
The authors gratefully acknowledge support of NSF Physics of Living Systems, under grant 1705139. The authors thank Dr. Sebastiaan Meenderink for developing the software used for tracking hair bundle movement.  

\subsection*{Author contributions statement}
D.B. conceived the experiments,  J.F. conducted the experiments, D.B. and J.F. analyzed the results.  D.B. and J.F. wrote the manuscript. 

\subsection*{Competing Interests}
The authors declare no competing interests.

\clearpage
\section*{Methods}

\subsection*{Analytic Approximation of the Lyapunov Exponent}

We use a similar approach to a previous Lyapunov exponent approximation.\cite{GOLDOBIN06}  Simulations show that the divergence of neighboring trajectories occurs predominantly in the $\hat{\theta}$ direction.  In the noiseless case, the $\hat{r}$ direction is stable, while the $\hat{\theta}$ direction is only marginally stable.  Therefore, we seek an approximation of the diverging perturbation in $\hat{\theta}$.  We look for an equation of the form $\frac{d  \langle \Delta\theta \rangle}{dt} = \lambda \langle \Delta\theta \rangle$, where $\lambda$ is the Lyapunov exponent.  Making the change of variables $z(t) = r(t)e^{i\theta(t)}$, equation (\ref{eq:Hopf}) becomes

\begin{ceqn}
\begin{align}
\frac{dr}{dt} = \mu r - \alpha r^3 + \eta_x(t)\cos\theta + \eta_y(t)\sin\theta
\end{align}
\end{ceqn}

\noindent and

\begin{ceqn}
\begin{align}
\frac{d\theta}{dt} = \omega_0 - \beta r^2 + \frac{1}{r}(\eta_y(t)\cos\theta - \eta_x(t)\sin\theta),
\end{align}
\end{ceqn}

\noindent where a nonzero $\beta$ yields nonisochronic dynamics.\cite{FABER19}  Now we express these two differential equations in terms of a small difference between two neighboring solutions ($r_1(t), \theta_1(t)$) and ($r_2(t), \theta_2(t)$).  We define $\Delta r = r_2 - r_1$  and $\Delta \theta = \theta_2 - \theta_1$.  Making this substitution yields

\begin{ceqn}
\begin{align}
\dot{\Delta r} = \mu \Delta r - \alpha ((r_1 + \Delta r)^3-r_1^3) + 2\sin(\frac{\Delta\theta}{2})\eta_1(t)
\label{eq:delta_rdot_full},
\end{align}
\end{ceqn}

\noindent where we have  defined new noise terms:

\begin{ceqn}
\begin{align}
\eta_1(t) = \cos(\frac{\theta_1 + \theta_2}{2}) \eta_y(t)  - \sin(\frac{\theta_1 + \theta_2}{2})\eta_x(t)
\end{align}
\end{ceqn}

\noindent and

\begin{ceqn}
\begin{align}
\eta_2(t) = \cos(\frac{\theta_1 + \theta_2}{2}) \eta_x(t)  + \sin(\frac{\theta_1 + \theta_2}{2})\eta_y(t)
\label{eq:1},
\end{align}
\end{ceqn}

\noindent which also have the properties $\langle \eta_1(t)\eta_1(t') \rangle  =  \langle \eta_2(t)\eta_2(t') \rangle = 2D \delta (t-t')$ and $\langle \eta_1(t)\eta_2(t') \rangle  = 0$.

Since the Lyapunov exponent is defined only in the limit of infinitesimal devations, we let $\frac{\Delta r}{r_1} << 1$, $\frac{\Delta r}{r_2} << 1$, and $\Delta \theta << 1$.  Keeping only the the first-order terms, equation (\ref{eq:delta_rdot_full}) becomes

\begin{ceqn}
\begin{align}
\dot{\Delta r} \approx \mu \Delta r - 3\alpha r_1^2 \Delta r + \Delta\theta\eta_1(t)
\label{eq:delta_rdot_2}.
\end{align}
\end{ceqn}

\noindent As the system spends the most time at the stable radius, we start one of the two solutions at this radius, $r_1 = r_0 = \sqrt{\frac{\mu}{\alpha}}$, and allow the second solution to be a perturbation from this radius, $r_2 = r_0 + \Delta r$.  Making this substitution, equation (\ref{eq:delta_rdot_2}) simplifies further:

\begin{ceqn}
\begin{align}
\dot{\Delta r} \approx -2\mu\Delta r + \Delta\theta \eta_1(t)
\label{eq:delta_rdot_3}.
\end{align}
\end{ceqn}

\noindent Notice that the dynamics are stable to perturbations in $r$.  However, as deviations in $\theta$ grow, so does the effective noise term, $\Delta\theta \eta_1(t)$, and trajectories will tend to spread farther away from the noiseless limit cycle radius.  We now use Fokker-Planck theory to find the probability distribution, $P(\Delta r)$, of this stable potential.  Inserting the drift and diffusion terms into the Fokker-Plank equation, we get

\begin{ceqn}
\begin{align}
\frac{\partial P}{\partial t} = - \frac{\partial}{\partial \Delta r} ( -2\mu \Delta r P ) + D(\frac{\partial}{\partial\Delta r})^2 ( (\Delta \theta)^2 P)
\label{eq:1}.
\end{align}
\end{ceqn}

\noindent We seek the steady-state solution, $\frac{\partial P}{\partial t} = 0$.

\begin{ceqn}
\begin{align}
2\mu \Delta r P  + D\frac{\partial}{\partial\Delta r} ( (\Delta \theta)^2 P) = constant = 0
\label{eq:1}.
\end{align}
\end{ceqn}

\noindent The constant must be zero in order for $P(\Delta r = \infty) = 0$.

\begin{ceqn}
\begin{align}
2\mu \Delta r P  + D(\Delta \theta)^2\frac{\partial P}{\partial\Delta r} + 2D(\Delta \theta) P\frac{\partial \Delta\theta}{\partial\Delta r}= 0
\label{eq:1}.
\end{align}
\end{ceqn}

We will assume that $P(\Delta \theta)$ reaches steady state quickly due to the stability of the limit cycle.  If the dynamics in $r$ can quickly stabilize upon variation in $\theta$, we can ignore the third term and easily find the probability distribution;

\begin{ceqn}
\begin{align}
\frac{\partial P}{\partial\Delta r} = - \frac{2\mu \Delta r}{D(\Delta \theta)^2} P
\label{eq:1}
\end{align}
\end{ceqn}

\begin{ceqn}
\begin{align}
P(\Delta r) = Ce^{-\frac{\mu \Delta r^2}{D(\Delta \theta)^2}}
\label{eq:p_distribution},
\end{align}
\end{ceqn}

\noindent where $C=\sqrt{\frac{\mu}{\pi D}}\frac{1}{\Delta\theta}$ is a normalization constant.  As expected, this distribution spreads out as we increase the noise strength or the angular deviation, $\Delta \theta$.  We now treat the $\dot{\theta}$ equation:

\begin{eqnarray}
\begin{split}
\dot{\theta_2} - \dot{\theta_1} = - \beta (r_2^2 - r_1^2) + \eta_y(t) (\frac{\cos\theta_2}{r_2} - \frac{\cos\theta_1}{r_1}) \\
- \eta_x(t) (\frac{\sin\theta_2}{r_2} - \frac{\sin\theta_1}{r_1}).
\label{eq:delta_thetadot_full}
\end{split}
\end{eqnarray}

\noindent Lyapunov exponents are calculated by averaging divergences rates over all of the phase space or, equivalently, over all time.  We therefore take the time average of equation (\ref{eq:delta_thetadot_full}) and only two terms survive:

\begin{ceqn}
\begin{align}
\langle \dot{\Delta \theta} \rangle = -2r_1\beta \langle\Delta r \rangle - \beta \langle (\Delta r)^2 \rangle .
\label{eq:1}
\end{align}
\end{ceqn}

\noindent We evaluate these average values using $P(\Delta r)$:

\begin{eqnarray}
\begin{split}
 \langle \dot{\Delta \theta} \rangle =  -2r_1\beta\int_{-\infty}^{\infty}\Delta r P(\Delta r) d\Delta r \\
 - \beta\int_{-\infty}^{\infty}(\Delta r)^2 P(\Delta r) d\Delta r
\label{eq:1}
\end{split}
\end{eqnarray}

\begin{ceqn}
\begin{align}
\langle \dot{\Delta \theta} \rangle = -\beta\sqrt{\frac{\mu}{\pi D}}\frac{1}{\langle \Delta\theta \rangle}\int_{-\infty}^{\infty}(\Delta r)^2 e^{-\frac{\mu \Delta r^2}{D(\Delta \theta)^2}} d\Delta r
\end{align}
\end{ceqn}

\begin{ceqn}
\begin{align}
\langle \dot{\Delta \theta} \rangle = \frac{d}{dt} \langle \Delta \theta \rangle = \frac{-\beta D \langle \Delta\theta \rangle^2}{2\mu}
\label{eq:1}
\end{align}
\end{ceqn}

\noindent This equation has a semi-stable fixed point at $\langle \Delta \theta \rangle = 0$.  In the presence of noise, this point is unstable.  Linearizing near the fixed point, we find that the solution diverges exponentially with Lyapunov exponent,

\begin{ceqn}
\begin{align}
\lambda = \frac{\lvert\beta\rvert D}{\mu}
\label{eq:LyapExp}.
\end{align}
\end{ceqn}

\noindent Using numerical simulations, we verify the validity of this approximation (see supplemental material, Fig. S1).

\subsection*{Transfer Entropy}

The transfer entropy\cite{SCHREIBER00} from process $J$ to process $I$ is defined as

\begin{ceqn}
\begin{align}
T_{J \rightarrow I} = \sum p(i_{n+1}, i_{n}^{(k)}, j_{n}^{(l)})
\log \frac {p(i_{n+1} \ | \ i_{n}^{(k)}, j_{n}^{(l)})} {p(i_{n+1} \ | \ i_{n}^{(k)})},
\label{eq:four}
\end{align}
\end{ceqn}

\noindent where $i_{n}^{(k)} = (i_n,...,i_{n-k+1})$ are the $k$ most recent states of process $I$.  Therefore, $p(i_{n+1} \ | \ i_{n}^{(k)}, j_{n}^{(l)})$ is the conditional probability of finding process $I$ in state $i_{n+1}$ at time $n+1$, given that the previous $k$ states of process $I$ were $i_{n}^{(k)}$ and given that the previous $l$ states of process $J$ were $j_{n}^{(l)}$.  The summation is performed over the length of the time series, as well as over all accessible states of processes $I$ and $J$.  Given the history of process $I$, the transfer entropy $T_{J \rightarrow I}$ is a measure of how much one's ability to predict the future of process $I$ is improved when one gains knowledge of the history of process $J$.  If these processes are completely unrelated, then $T_{J \rightarrow I} = 0.$  We discretize the recordings of hair bundle position into two bins, a natural choice due to the bimodal distribution in position of the hair bundle.  Likewise, the bimodal burst noise stimulus is characterized by two states.  The choice of $k$ and $l$ has little effect on our results, so we select $k = l = 5$.

\bigbreak
\bigbreak

\subsection*{Experimental Techniques}

\bigbreak

\subsubsection*{Biological Preparation}
Experiments were performed \textit{in vitro} on hair cells of the American bullfrog (\textit{Rana catesbeiana}) sacculus, an organ responsible for low-frequency air-borne and ground-borne vibrations.  Sacculi were excised from the inner ear of the animal, and mounted in a two-compartment chamber with artificial perilymph and endolymph solutions.\cite{BENSER96}  Hair bundles were accessed after digestion and removal of the overlying otolithic membrane.\cite{MARTIN01}  All protocols for animal care and euthanasia were approved by the UCLA Chancellor's Animal Research Committee in accordance with federal and state regulations.  

\subsubsection*{Mechanical Stimulus}
To deliver a stimulus to the hair bundles, we used glass capillaries that had been melted and stretched with a micropipette puller.  These elastic probes were calibrated by observing their Brownian motion with a high-speed camera and applying the fluctuation dissipation theorem.  Typical stiffness and drag coefficients of these probes were 50 -- 150 $\mu N/m$ and 100 -- 200 $nNs/m$, respectively.  These elastic probes were treated with a charged polymer that improves adhesion to the hair bundle.  Innate oscillations persisted after the attachment of a probe.  The position of the probe base was controlled with a piezoelectric actuator.  Stimulus waveforms were delivered to the actuator using LabVIEW.

\subsubsection*{Data Collection}
Hair bundle motion was recorded with a high-speed camera at framerates of 500 Hz or 1 kHz.  The records were analyzed in MATLAB, using a center-of-pixel-intensity technique to determine the position of the center of the hair bundle in each frame.  Typical noise floors of this technique, combined with stochastic fluctuations of bundle position in the fluid, were 3 -- 5 nm.

\subsubsection*{Stimulus Waveforms}
Experiments were carried out as follows.  First, we obtained a 60 second recording of the spontaneous oscillation, immediately followed by sinusoidal stimuli applied at several frequencies (20 stimulus cycles for each frequency).  Then, the hair bundle was stimulated with burst noise (random telegraph noise), which was generated by randomly selecting time intervals between rising and falling flanks of the square wave.  The intervals were selected such that the frequencies of the square waves ranged from 3 to 50 Hz, all with equal probability.  This distribution spans the full frequency range of typical spontaneously oscillating hair bundles in the American Bullfrog sacculus\cite{MARTIN03} and is comparable to the 4-octave range of stimulus in our simulations.  The wide range and flat probability distribution ensured that a change in sensitivity could not be due to a simple shift in the natural frequency of spontaneous oscillations.  This stimulus lasted 20 seconds and included 300-400 full square waves.  After this initial recording, the experimental parameters (calcium concentration and viscosity of the endolymph) were varied, and identical stimulus protocols were delivered again. Recordings were obtained under several different variations of calcium concentration and/or viscosity, so as to elicit different degrees of chaos from the same hair cell.

\subsubsection*{Data Analysis}
The Kolmogorov entropy was calculated from the 60 second recording segment with no stimulus.  We scaled these measurements to the time scale of each recording, $\tau$, which was taken to be the time for the autocorrelation function of the spontaneous oscillations to cross zero.  This ensured that a change in K-entropy was indeed a change in the predictability of the system and not simply a shift in the natural frequency.  A consequence of this scaling was that it rendered the noise floor on K-entropy ($\sim 0.5$ bits/$\tau$) large with respect to the measurements ($0.5 - 1.3$ bits/$\tau$).  

The response to sinusoidal stimulus and the spontaneous oscillations were used to calculate the gain.  The response to the burst noise was used to calculate the transfer entropy, the mean displacement, and the response time, using the same techniques as described in the Theoretical Results.

Bootstrap data sets were generated by adding fluctuations to the original data set based on the uncertainty in measuring hair bundle position.  The position measurement uncertainty was quantified by recording a stationary object with the high-speed camera.  To generate a bootstrap data set, each measurement of position in the original data set was given a random perturbation of magnitude based on the statistics of the fluctuations in position of the stationary object.

\end{multicols}
\end{document}


\title{Supplemental Material for: Chaotic Dynamics Enhance the Sensitivity of Inner Ear Hair Cells}
\author{Justin Faber$^1$}
\author{Dolores Bozovic$^{1, 2}$}
\affiliation{ $^1$Department of Physics \& Astronomy and $^2$California NanoSystems Institute, University of California, Los Angeles, California 90095, USA\\}
\date{\today}

\maketitle

\begin{figure*}[h]
\includegraphics[width=\textwidth]{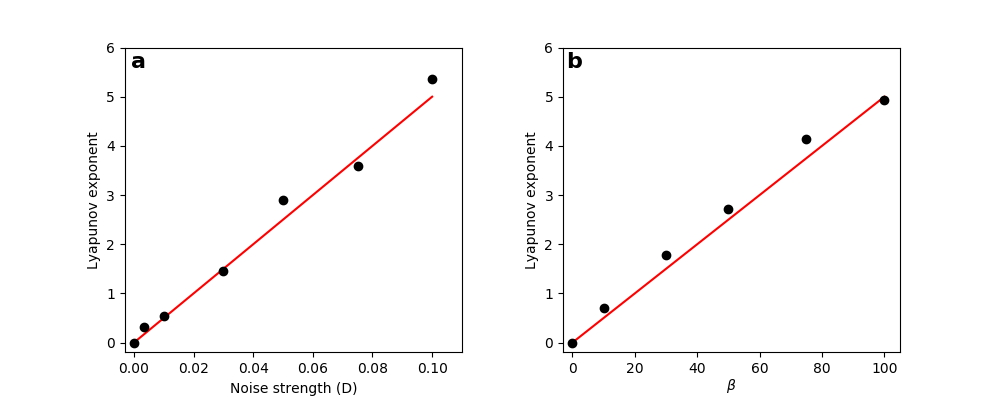}
\caption{Comparison of the analytic calculation of the Lyapunov exponent (red line) to the numerical calculation (black points).  (\textbf{a}) The noise strength is varied, while all other parameters are fixed ($\mu = \alpha = \Omega_0 = 1$, $\beta = 50$).  (\textbf{b})  $\beta$ is varied, while all other parameters are fixed ($\mu = \alpha = \Omega_0 = 1$, $D = 0.05$)}
\end{figure*}

\begin{figure*}[t!]
\includegraphics[width=4in]{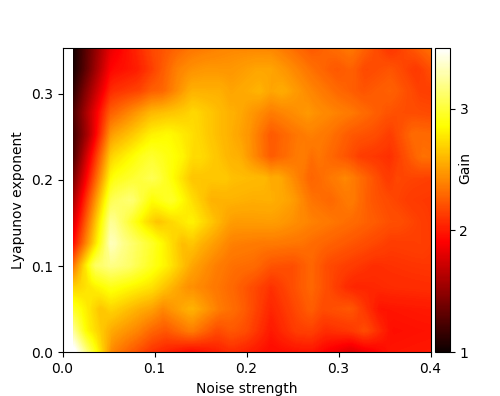}
\caption{Phase-locked amplitude gain for above-resonance ($\omega_{stim} = 1.05\Omega_0$), sinusoidal stimulus as the noise strength and Lyapunov exponent are varied.  Color was generated by linearly interpolating a grid of 10 Lyapunov exponent values and 10 noise strengths.}
\end{figure*}

\begin{figure*}[t!]
\includegraphics[width=\textwidth]{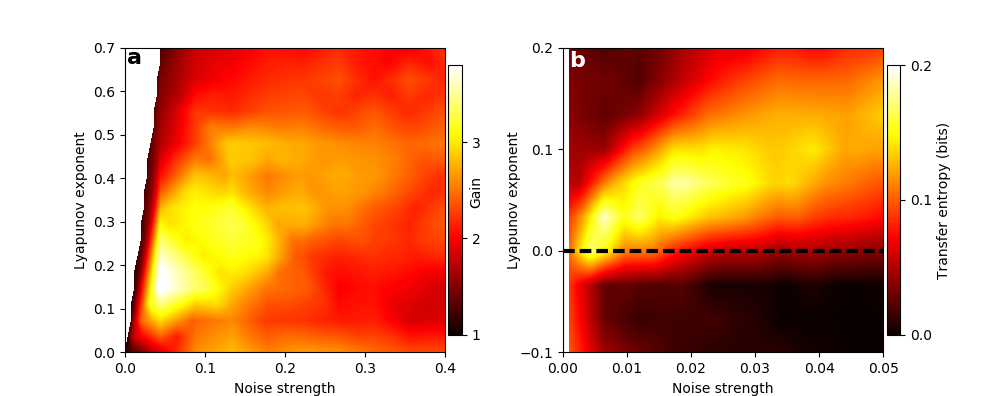}
\caption{(\textbf{a}) Phase-locked amplitude gain for on-resonance, sinusoidal stimulus as the noise strength and Lyapunov exponent are varied.  Color was generated by linearly interpolating a grid of 10 Lyapunov exponent values and 10 noise strengths.  (\textbf{b})  Transfer entropy from burst noise stimulus to response of the Hopf oscillator as noise strength and Lyapunov exponent are varied.  In the $\lambda < 0$ regime, the system is quiescent, and the Lyapunov exponent characterizes the stability of this fixed point.  Color was generated by linearly interpolating a grid of 10 Lyapunov exponent values and 10 noise strengths.  For both panels, the Lyapunov exponent was modulated  by varying $\mu$.}
\end{figure*}

\begin{figure*}[t!]
\includegraphics[width=4in]{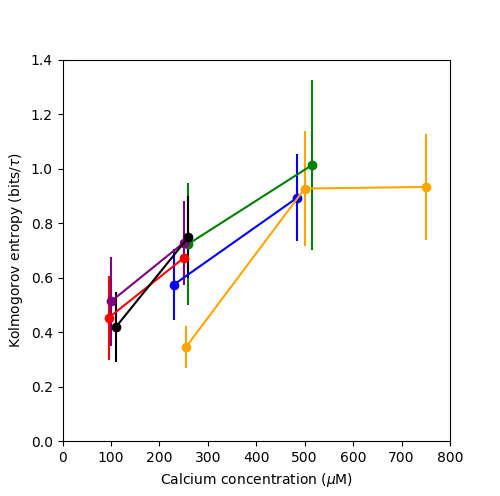}
\caption{The dependence of Kolmogorov entropy on the endolymph calcium concentration.  Each of the six colors corresponds to a different cell.}
\end{figure*}

\begin{figure*}[t!]
\includegraphics[width=\textwidth]{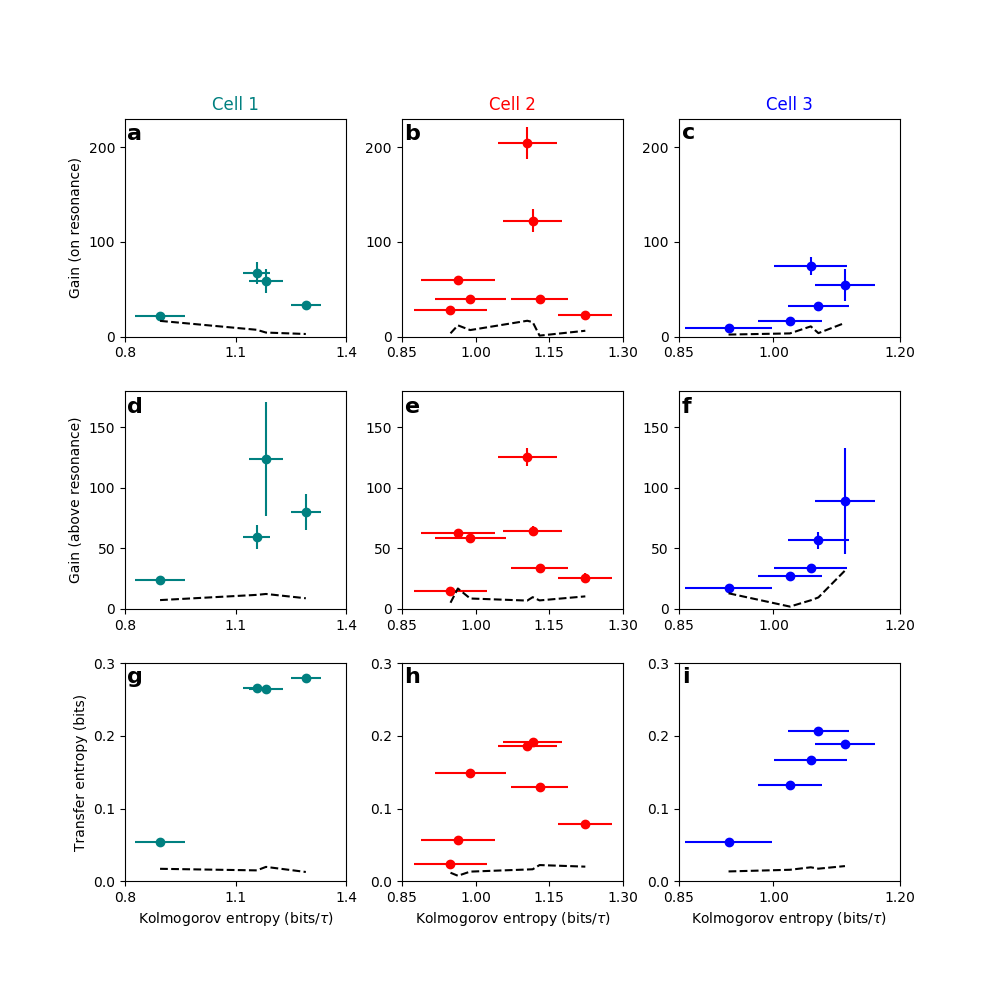}
\caption{Phase-locked amplitude gain for 2 pN sinusoidal stimulus at the natural frequency (\textbf{a-c}) and above the natural frequency (\textbf{d-f}) for three additional cells.  Data points and error bars on the gain represent the mean and standard deviation from 100 bootstraps. The noise floor (dashed curve) was calculated by treating a segment of the spontaneous oscillation recording as if a stimulus were present and calculating the gain.  This curve represents the mean plus one standard deviation from 100 bootstraps.  (\textbf{g-i}) Transfer entropy from burst noise stimulus to response.  Data points and error bars represent the mean and standard deviation obtained from 100 bootstraps.  The noise floor (dashed curve) was determined by calculating the transfer entropy in the reverse direction (response to stimulus).  This curve represents the mean plus one standard deviation from 100 bootstraps.}
\end{figure*}

\begin{figure*}[t!]
\includegraphics[width=\textwidth]{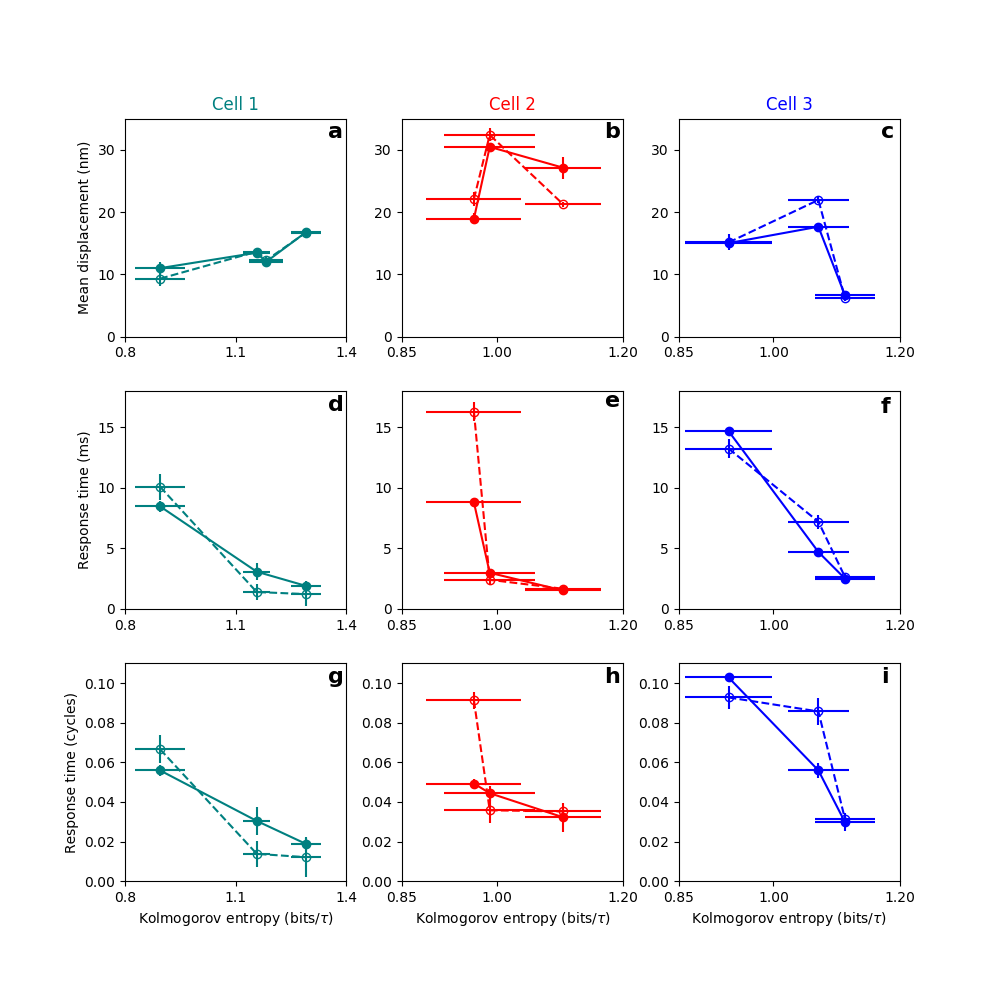}
\caption{(\textbf{a-c})  Average displacement induced on the hair bundle from the step stimulus, averaged over $\sim200$ square waves.  Data points and error bars represent the mean and standard deviation of the response plateau.  (\textbf{d-f})  Response time to step stimulus, characterized by fitting the mean response to an exponential and taking the decay time.  Error bars represent the standard deviation of the residual associated with the exponential fit.  (\textbf{g-i}) Response times from (d-f) scaled to the natural periods of oscillation for each recording.  For all panels, open-dashed and closed-solid data points represent averages over steps in the positive (channel open) and negative (channel closed) directions, respectively.}
\end{figure*}